# Dielectric Metamaterials with Toroidal Dipolar Response

Alexey A. Basharin[1,2], Maria Kafesaki[1,3], Eleftherios N. Economou[1,4] and Costas M. Soukoulis[1,5]

[1] *Institute of Electronic Structure and Laser (IESL), Foundation for Research and Technology Hellas (FORTH), P.O. Box 1385, 71110 Heraklion, Crete, Greece*
[2] *National Research University "Moscow Power Engineering Institute", 112250 Moscow, Russia*
[3] *Department of Materials Science and Technology, University of Crete, 71003 Heraklion, Greece*
[4] *Department of Physics, University of Crete, 71003 Heraklion, Greece*
[5] *Ames Laboratory-USDOE, and Department of Physics and Astronomy, Iowa State University, Ames, Iowa 50011, USA*

Vassili A. Fedotov[6], Vassili Savinov[6] and Nikolay I. Zheludev[6,7]

[6] *Optoelectronics Research Centre and Centre for Photonic Metamaterials, University of Southampton, SO17 1BJ, UK*
[7] *Centre for Disruptive Photonic Technologies, Nanyang Technological University, 637371, Singapore*

*basharin@iesl.forth.gr*

**Abstract: Toroidal multipoles are the terms missing in the standard multipole expansion; they are usually overlooked due to their relatively weak coupling to the electromagnetic fields. Here we propose and theoretically study all-dielectric metamaterials of a special class that represent a simple electromagnetic system supporting toroidal dipolar excitations in the THz part of the spectrum. We show that resonant transmission and reflection of such metamaterials is dominated by toroidal dipole scattering, the neglect of which would result in a misunderstanding interpretation of the metamaterials' macroscopic response. Due to the unique field configuration of the toroidal mode the proposed metamaterials could serve as a platform for sensing, or enhancement of light absorption and optical nonlinearities.**

Toroidal dipole is a peculiar elementary *current* excitation corresponding to electrical currents circulating on a surface of a gedanken torus along its meridians (so-called poloidal currents). Toroidal dipole was first considered in 1957 by Zel'dovich, who suggested that such excitation produced by static currents (anapole), if supported by atomic nuclei, could explain parity violation in the weak interaction [1]. Since then the existence of the static toroidal dipole has been predicted and its importance has been discussed for a number of solid-state systems including ferroelectric and ferromagnetic nano- and micro-structures, multiferroics, macromolecules, molecular magnets etc [2-9]. The dynamic toroidal dipole is less known. Although it was shown to radiate electromagnetic fields, just as the conventional dynamic multipoles do, toroidal dipole is not included in the standard multipole expansion and thus often omitted from considerations in classical electrodynamics [10, 11]. Being physically distinct from the dynamic electric dipole (which is produced by oscillating charge density rather than currents), the toroidal dipole emits radiation with the same angular momentum and parity properties, which is therefore indistinguishable from that of an electric dipole for any distant observer [12]. There are, however, fundamental differences between the two types of radiation: the radiated power for electric and toroidal dipoles scales as respectively $\omega^4$ and $\omega^6$ [10, 11, 13]; also the vector-potential fields corresponding to the electric and toroidal dipolar emission do not coincide and the difference cannot be removed by gauge fixing [12].

Other intriguing phenomena expected in the presence of toroidal dipolar excitations include the violation of the action-reaction equality [14], non-reciprocal refraction of light [15], and the generation of propagating non-trivial vector potential in the complete absence of electromagnetic fields [12]. The existence of the dynamic toroidal dipole also indicates that care should be taken while establishing a relation between the far-field properties of an electromagnetic source and the charge-current distribution forming the source. This applies to many domains of science dealing with electromagnetic interactions, and in particular for nanophotonics and plasmonics where the topology of charge-current excitations controls the enhancement of local optical fields [16-24]. Furthermore, given the explicit toroidal topology of a great number of biologically important macromolecules and protein complexes[25-27], it is not unreasonable to expect that electromagnetic interactions involving toroidal dipole (and higher toroidal multipoles) could play a special role in nature.

The detection of toroidal excitations is challenging [28]. The dynamic toroidal dipole interacts with curl **B** and is weakly coupled to free space, while its manifestation can be naturally masked by much stronger electromagnetic effects due to electric and magnetic dipoles and even electric quadrupole. Experimental observation of the toroidal response has become possible only recently with use of the metamaterial concept [29]. This concept enables to access to novel and exotic optical phenomena (such as negative refraction and cloaking [30-36]) by controlling the symmetry and character of electromagnetic response through artificially structuring media on a sub-wavelength scale. The toroidal dipolar response was demonstrated in a metamaterial composed of specially designed complex metallic electromagnetic scatterers (metamolecules) of toroidal topology, where electric and magnetic dipolar excitations had been suppressed, while toroidal response was spectrally isolated and resonantly enhanced to a detectable level [29]. This demonstration has opened the way towards testing the astonishing predictions of the toroidal electrodynamics [37] and stimulated works on developing metamaterial and plasmonic systems exhibiting strong toroidal response [38-45].

In this paper we aim to eliminate one of the major drawbacks of the toroidal metamaterial designs reported so far, namely the dissipation loss. It results from the Ohmic resistance of the metallic resonators exploited by the existing designs and can substantially hinder the excitation of toroidal multipoles given their weak coupling to the external fields. To address this problem we propose a completely non-metallic (all-dielectric) metamaterial that is virtually free from dissipation and is capable of supporting strong resonant toroidal excitations in the terahertz part of the spectrum. We also show that owing to the unique topology of the toroidal dipolar mode its electric field can be spatially confined within sub-wavelength externally accessible dielectric-free regions of the metamolecules, which makes the proposed metamaterial a viable platform for sensing, and enhancement of light absorption and optical nonlinearities.

Our dielectric metamaterial is based on clusters of sub-wavelength high-index dielectric cylinders. Each cluster contains 4 cylinders and represents the elementary building block of the metamaterial, i.e. its metamolecule (see Fig. 1).

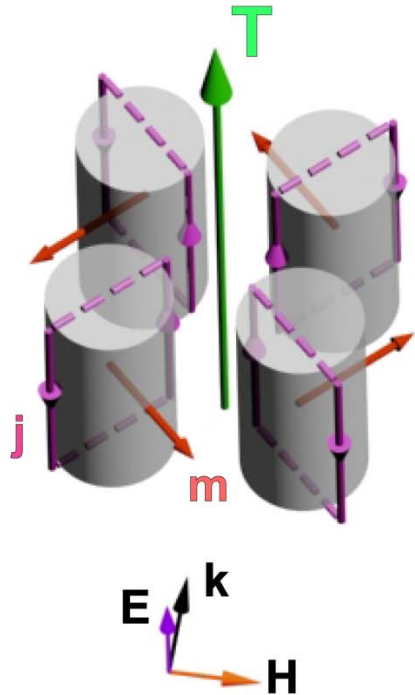

Figure 1. A fragment of modeled dielectric metamolecule supporting toroidal dipolar excitation. The metamolecule is composed of four infinitely long closely spaced high-index dielectric cylinders. **j** - displacement currents induced by vertically polarized (i.e. along the cylinders axes) plane wave, **m** - magnetic dipole moments of the constituent dielectric cylinders, **T** - net toroidal dipole moment of the metamolecule.

As of any dielectric structure, the electromagnetic response of such a metamolecule is underpinned by the displacement rather than conduction currents, which are induced in each cylinder by an incident electromagnetic wave. Depending on the radius and length of the cylinders, their dielectric constant and the polarization of the incident wave, the electromagnetic scattering produced by the displacement currents can become resonant. It is this resonance (known as the Mie resonance [46]) that in the dielectric metamaterials is the counterpart of the lossy plasmonic resonance characteristic to the metallic metamolecules [47-52]. It was shown, in particular, that the Mie resonances of individual high-index dielectric cylinders could lead to resonant magnetic dipolar response for orthogonal [53] as well as parallel [54] polarizations of the incident wave, thus rendering the dielectric metamaterials as a lossless solution for engineering the artificial magnetism [55].

The challenge of engineering toroidal response is to create a metamolecule that would support dynamically induced and spatially confined magnetization circulating along a loop. In our case such metamolecule is formed by placing 4 dielectric cylinders close to each other, thereby promoting near-field coupling between the Mie-type magnetic modes excited individually in each cylinder. These modes correspond to the displacement currents **j** oscillating in the inward and outward parts of the cylinders in the opposite directions and are excited by the plane electromagnetic wave with **E**-vector parallel to the axes of the cylinders (see Fig. 1). We will show below that for a narrow range of frequencies the magnetic moments of the modes **m**, which oscillate perpendicular to the axes of the cylinders, may all become aligned head-to-tail forming a dynamic vortex state with closed loops of the magnetic field confined well within the metamolecule. In an ideal case such state is characterized by zero net magnetic and electric multipole moments, and a non-zero toroidal dipole moment oscillating along the axis of the metamolecule (see Fig. 1).

We assume in our model that the cylinders are made of LiTaO$_3$, an ionic crystal that is known to exhibit strong polaritonic response at THz frequencies due to the excitation of optical phonons [48, 56-58]. LiTaO$_3$ cylinders can be practically realized using various methods of crystal growth.[57, 59] The complex dielectric permittivity of LiTaO$_3$ displays dispersion of the Lorenz type:

$$\varepsilon = \varepsilon_\infty \frac{\omega^2 - \omega_L^2 + i\omega\gamma}{\omega^2 - \omega_T^2 + i\omega\gamma}, \tag{1}$$

where $\omega_T/2\pi$ = 26.7 THz is the frequency of the transverse optical phonons, $\omega_L/2\pi$ = 46.9 THz is the frequency of longitudinal optical phonons, $\gamma/2\pi$ = 0.94 THz is the damping factor due to dipole relaxation and $\varepsilon_\infty$ = 13.4 is the limiting value of the permittivity for frequencies much higher than $\omega_L$. At frequencies well below the phonon resonances the dielectric permittivity of LiTaO$_3$ can be rather high reaching $\varepsilon$ = 41.4. Importantly, at these frequencies the crystal exhibits negligible dissipation loss, which makes it a standout candidate among other polaritonic materials. The LiTaO$_3$ cylinders have the radius $R = 8$ μm and are arranged inside the metamolecular cluster with the centre-to-centre separation $a=18$ μm. The clusters are surrounded by vacuum/air and periodically placed along the *x*-axis with period $d = 58$ μm. The cylinders are also assumed to be infinitely long, which allows us to describe the resulting metamatrerial slab using a two-dimensional model (see Fig. 2a). The electromagnetic properties of the resulting metamaterial slab were modeled using the commercial Maxwell's equations solver CST Microwave Studio.

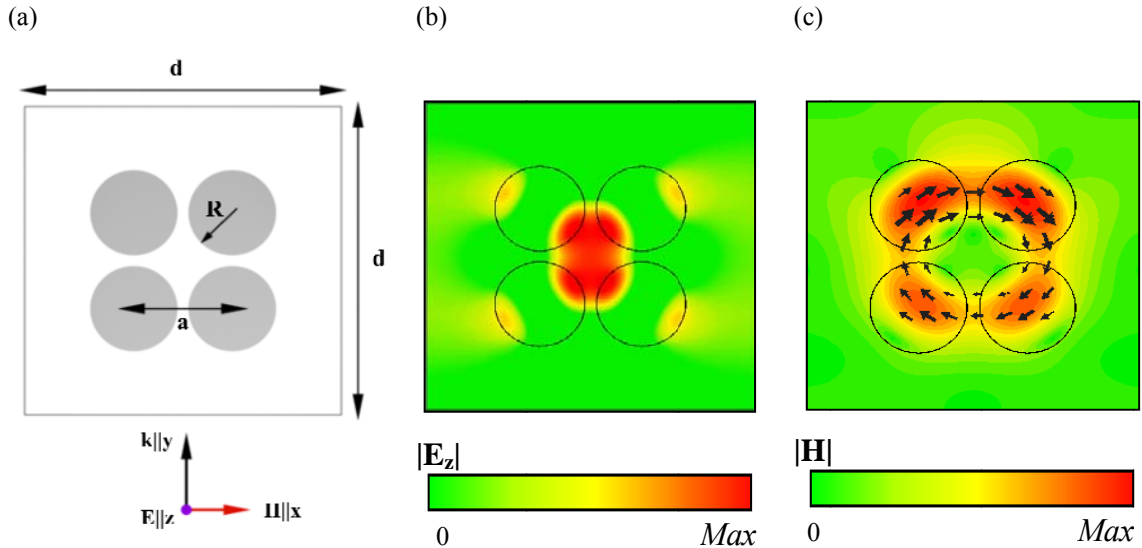

Figure 2. (a) Cross-section of a dielectric metamolecule supporting resonant toroidal dipolar excitation for the incident electromagnetic field configuration shown at the bottom of the panel. The metamolecule is a cluster of 4 infinitely long LiTaO$_3$ cylinders of $R = 8$ μm arranged with center-to-center separation $a = 18$ μm. The metamaterial is formed by placing the metamolecules periodically along the *x*-axis with period $d = 58$ μm. (b) & (c) Calculated distributions of the absolute values of electric field (*z*-component) (panel b) and magnetic field (panel c) induced in the metamolecule at 1.89 THz. The slight asymmetry of the magnetic-field distribution is attributed to the presence of non-resonant magnetic dipolar mode supported by the metamolecule (see Fig. 3b).

The toroidal dipolar response was observed at around 1.89 THz as a resonance of full transmission (see Fig. 3a). The appearance of the toroidal excitation here was confirmed by calculating the distribution of the local magnetic field induced in the metamolecule, which we present in Fig. 2c. The field map clearly shows a distinct vortex of the magnetic field that threads all four cylinders of the metamolecule. Although the appearance of a magnetic vortex is also expected for an electric dipolar excitation (think of magnetic field lines circling around electrical current), the induced magnetic field in the present case is confined within a well-defined ring-like region ('torus'). Such spatial localization of the magnetic field is the key signature of the toroidal dipole mode.

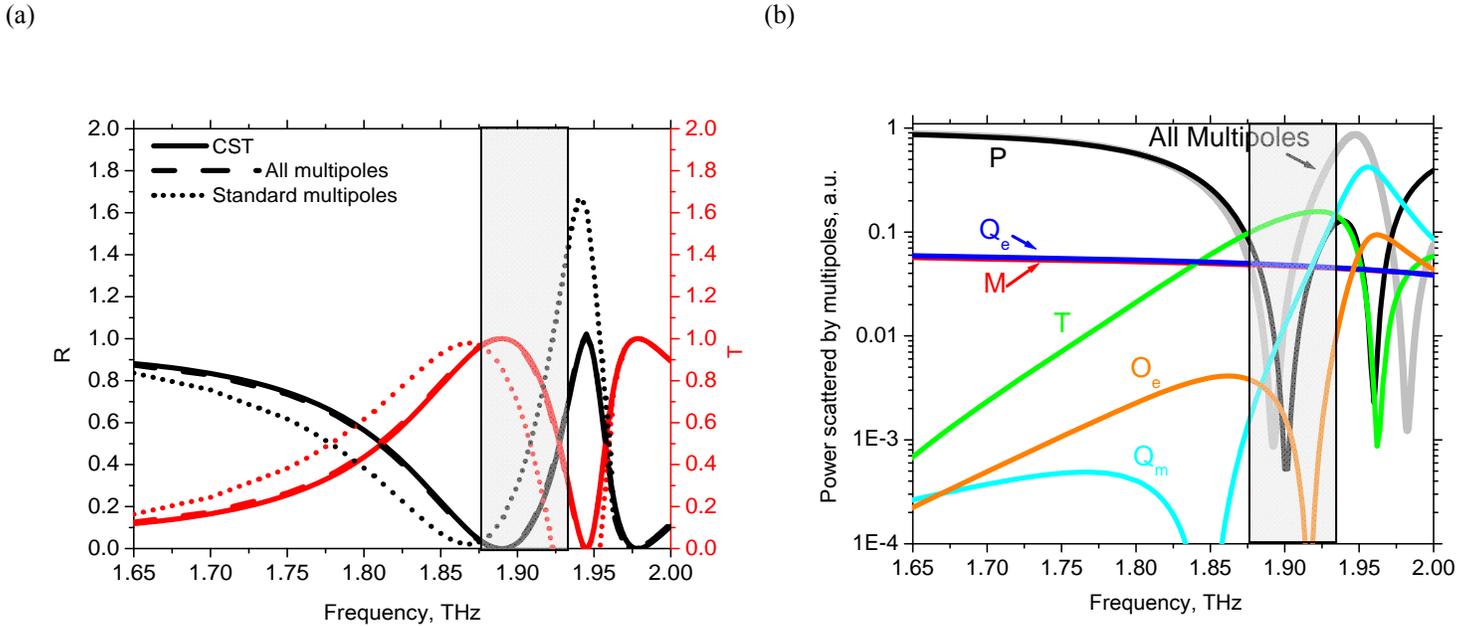

Figure 3. (a) Transmission, T (red online), and reflection, R (black), spectra calculated for the metamaterial slab composed of 4-cylinder clusters periodically placed as described in Fig. 2. Solid curves correspond to the results of CST Microwave Studio simulations. Dashed curves were obtained based on multipole scattering data, which included all multipole terms. Dotted curves were obtained based on multipole scattering data, which included only standard multipole terms, avoiding the torroidal dipolar contribution. (b) Individual contributions of the six strongest multipolar excitations to the power radiated (scattered) by our metamaterial array near its resonance.

To assess the role of the toroidal excitation in forming the observed resonant response we compared the relative contributions [60] of the toroidal dipole and standard multipoles to backward and forward scattering produced by the metamaterial slab. The multipole moments induced in the metamolecules were calculated based on the density of the displacement currents, i.e. currents resulted from the displacement of charges in the dielectric. The latter were extracted from the simulated near-field distribution data using the following formula [61]

$$\mathbf{j} = \frac{d\mathbf{P}}{dt} = \varepsilon_0(\varepsilon-1)\frac{d\mathbf{E}}{dt} = i\omega\varepsilon_0(\varepsilon-1)\mathbf{E}. \qquad (2)$$

This approach allowed us to unambiguously link the near-field signature of the multipolar excitations to their electromagnetic response in the far field. The latter would not be possible with the multipole expansion based on radiated fields (as featured, for example, in the Mie-type scattering analysis) since the same-order toroidal and electric multipoles emit radiation with the same angular momentum and parity properties and thus appear indistinguishable to any *distant observer* [12].

The multipolar scattering was characterized in terms of the power of electromagnetic radiation re-emitted in the form of plane waves by the arrays of induced multipoles [62]. For each type of the multipoles the radiated field was calculated by summing the corresponding multipolar contributions of the individual metamolecules at a point located in the far-field zone. All the metamolecules were assumed to be excited in phase as expected in the case of plane-wave illumination at normal incidence; the radiation patterns of the multipolar excitations were described through the spherical vector harmonics (see [60] for more details). The results of such calculations are presented in Fig. 3b, which shows the power radiated by the arrays of induced toroidal dipole **T** and the five strongest multipoles, namely electric dipole **P**, magnetic dipole **M**, electric quadrupole $\mathbf{Q_e}$, magnetic quadrupole $\mathbf{Q_m}$ and electric octupole $\mathbf{O_e}$. One can see that at the resonance the contribution of the electric dipole is strongly suppressed and there is a narrow range of frequencies, from 1.87 to 1.93 THz (shown by the gray box in Fig. 3) where the far-field scattering due to the resonant toroidal excitation dominates all other standard multipoles. In particular, the power scattered through the toroidal dipole moment in this frequency range is more than two times higher than that of the magnetic dipole, and electric and magnetic quadrupoles moments, and more than 25 times higher than the power scattered through the electric octupole moment. The key role of the toroidal excitation can be further demonstrated by calculating reflection, R, and transmission, T, of the metamaterial slab using the obtained multipole scattering data. In particular, reflection is given by the total power scattered by the metamolecules in the backward direction; transmission is generally given by the coherent superposition of the incident and forward-scattered fields but in the absence of losses (our case) it can be calculated simply as T = 1 – R. We stress from Fig. 3a that the metamaterial response can be exactly reproduced by taking into account all the multipole contributions, whilst ignoring the toroidal dipole moment makes the correct replication of the observed resonance impossible. Furthermore, failure to account for the toroidal dipole moment leads to simply unphysical behavior of the metamaterial response near 1.95 THz where its transmission exceeds 100 % (see dotted curve in Fig. 3a). At higher frequencies though the discrepancy in the calculated spectra appears to be much smaller, which is the result of toroidal dipole scattering contribution becoming negligible.

We note in passing that the resonant excitation of the toroidal dipole mode results from the coupling of the magnetic Mie modes of the four cylinders comprising our metamolecule. The frequency of the first "magnetic" mode of a single dielectric cylinder, which in our case is 2.18 THz, can be estimated analytically by setting to zero the denominator of the Mie scattering coefficient[63]

$$b_1 = \frac{nJ_1'(z_2)J_1(z_1) - J_1(z_2)J_1'(z_1)}{nJ_1'(z_2)H_1^{(1)}(z_1) - J_1(z_2)H_1^{(1)'}(z_1)}, \quad z_2 = k_2 R, \quad z_1 = k_0 R, \quad (3)$$

where $k_0 = \omega/c$ is the wave vector in free space, $k_2 = k_0\sqrt{\varepsilon}$, $n = \sqrt{\varepsilon}$, and $J_1$ and $H_1^{(1)}$ the Bessel function and the first-kind Hankel function of first order, respectively. The lowering of the toroidal resonance frequency compared to the Mie-mode frequency is the direct consequence of the near-field coupling between the magnetic Mie modes induced in each cylinder, which in the case of the circular head-to-tail dipole arrangement (see Fig. 1) reduces the energy of the resulting toroidal mode.

Importantly, the mode's electric field has only the longitudinal component (parallel to the axis of the cylinders) and is concentrated in the central part of the metamolecule in a spot smaller that λ/5 (see Fig. 2b). Such localization of the electric field is characteristic only to the toroidal dipolar mode, where the electric field is laterally confined within the eye of the sub-wavelength magnetic-field vortex and is sustained there by **curl H** in the absence of polarization or charge-density oscillations. As a result, the field distribution is seen to reach maximum in the dielectric-free region of the metamolecule rendering it as a sub-wavelength externally accessible cavity that can be exploited, for example, in sensing or achieving strong optical absorption or nonlinear response. In particular, such cavity may be seen as a part of a microfluidic system designed for high-throughput biological screening, or biomedical and environmental monitoring in the THz part of the spectrum. Strongly localized electric field of the toroidal mode will also enhance electronic nonlinearities of semiconductors, polaritonic nonlinearities of ferroelectrics and THz Kerr effect in liquids placed in the central area of the metamolecules, which provides a platform for metamaterial-assisted nonlinear THz-spectroscopy and second-harmonic generation [64].

Apart from engaging the toroidal dipolar mode, the localization of electric field outside the dielectric region can be achieved in the gap between two transversally polarized cylinders. In this case, however, the externally applied electric field must be non-resonant (quasi-static), since the fields of the resonantly induced Mie modes are localized inside the cylinders. Such a capacitor-like configuration provides confinement of the electric field in only one dimension, while obtaining large field enhancement in the gap would require dielectrics with the permittivity substantially higher than 41.4 .

Finally, we would like to comment on the mechanism of the metamaterial's electromagnetic transparency observed near the toroidal resonance (Fig. 3a). It results from the interference of the fields scattered by non-resonant magnetic dipolar and electric quadrupolar modes, and by resonant toroidal dipolar mode (see Fig. 3b). Although the angular distributions of the multipolar scattering are anisotropic and normally do not match for the individual multipoles of different types (except for the electric and toroidal multipoles [11, 13]), the omni-directional cancellation of the fields scattered by $\mathbf{T}$, $\mathbf{M}$ and $\mathbf{Q_e}$ is nevertheless possible for the arrays of these multipoles arranged on a regular grid with sub-wavelength periodicity (as it happens in the metamaterial slab). Indeed, the emission diagrams of such multipole arrays collapse around the normal to their plane, signifying the regime of plane-wave scattering with only two scattering directions available - forward and backward. In our case, the plane waves re-emitted by the combination of induced magnetic dipoles and electric quadrupoles oscillate in phase but with a π-delay compared to the waves produced by the toroidal dipoles. This makes their far-field interference destructive in both forward and backward directions, which is complete at around 1.89 THz. We believe that the resonant transparency of the proposed dielectric metamaterials may provide an alternative route towards achieving the non-trivial non-radiating charge-current excitation which was originally shown to exist for an isolated toroidal dipole destructively interfering with an electric dipole [12].

In conclusion, we proposed and theoretically studied a novel class of all-dielectric metamaterials that exhibit resonant toroidal dipolar response in the THz part of the spectrum. Our metamaterials are based on sub-wavelength clusters of high-index dielectric cylinders operating in the regime of resonant Mie scattering. We show that the near-field coupling between the individual Mie modes of the cylinders is capable of suppressing all standard multipoles, and make electromagnetic scattering due to the resulting toroidal dipolar excitations the dominant mechanism of the metamaterial response. Our findings thus indicate that the notion of the toroidal dipole can be crucial for the correct interpretation of electromagnetic properties in artificially structured material systems. The proposed metamaterials can be readily fabricated from low-loss polaritonic material $LiTaO_3$ and, given the unique topology of the toroidal dipolar mode, may be employed as a platform for sensing or enhancement of light absorption and optical nonlinearities.

ACKNOWLEDGEMENTS

Authors acknowledge the financial support by the EU projects ENSEMBLE and By-Nanoera and the Greek project ERC-02 ExEL (grant no. 6260). Work at Ames Laboratory was partially supported by the Department of Energy (Basic Energy Sciences) under Contract No. DE-AC02-07CH11358 (computational studies). Work at National Research University "Moscow Power Engineering Institute" was supported by RFBR (grant agreement no. 13-02-00732 and no.13-08-01278) and Russian Federal Program "Scientific and Scientific-Pedagogical Staff of Innovative Russia" for 2009-2013 (the Agreement no. 14.B37.21.1211). Authors at the University of Southampton acknowledge financial support of the Royal Society and the UK's Engineering and Physical Sciences Research Council under the Programme Grant on Photonic Metamaterials and Career Acceleration Fellowship (V. A. F.). Work at Nanyang Technological University was partially supported by the MOE Singapore grant MOE2011-T3-1-005.

APPENDIX

We also characterized the spectral response of the proposed dielectric metamaterial with respect to the angle of incidence. Figure A1a shows the results of our simulations for the case of TE polarization where the electric vector of the plane incident wave was kept parallel to the axes of the cylinders. Evidently, the toroidal resonance can be observed at oblique incidence and it is seen to shift to higher frequencies with increasing angle due spatial dispersion of the metamaterial slab own to the not extreme subwavelength size of the unit cell. The persistence of the toroidal mode was also confirmed by calculating the corresponding magnetic near-field distribution, as well as by comparing its contribution to the far-field scattering with that of the leading standard dipoles (see Figs. A1b and A1c). Although at large angles of incidence the ring-like pattern of the mode's magnetic field appears to be disturbed (due to the modified interference between the toroidal mode and the incident field) , it still features a pronounced anti-symmetric component that ensures strong toroidal dipolar response for each metamolecule (see Fig. A1b).

(a)

(b)

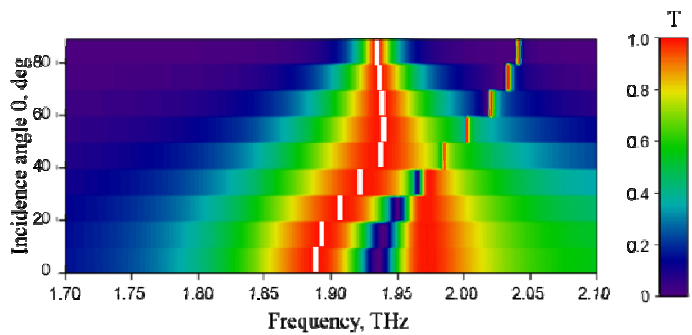
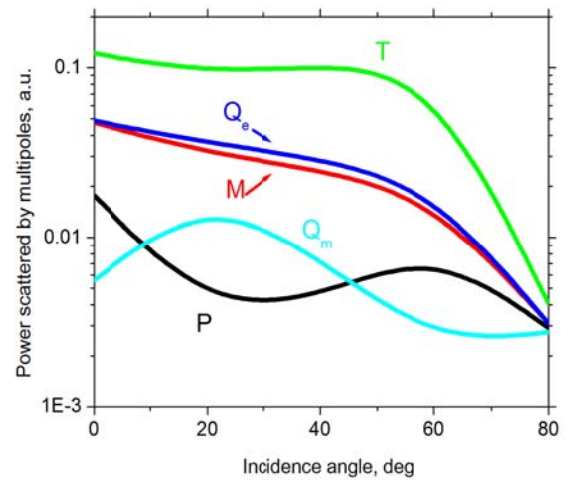

(c)

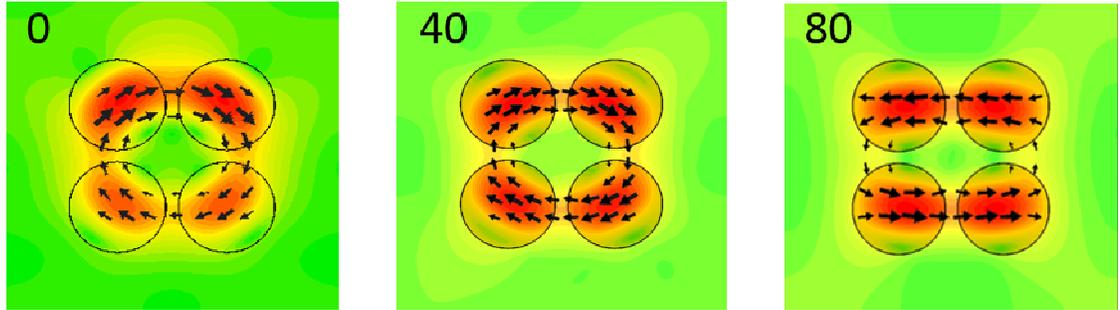

Figure A1. (a) Transmission, T, spectra of the metamaterial slab calculated with CST Microwave Studio for obliquely incident *E*-polarized plane waves (electric vector of the waves was set parallel to the axes of the cylinders) with the step of 10 degrees. White bars mark the frequency of the toroidal mode. (b) Relative strength of four leading standard multiples and toroidal dipole compared for various angles of incidence at the frequency of the toroidal mode. (c) Magnetic near-filed distributions calculated for angles of incidence $\theta = 0°$, 40°, 80° at the frequency of the toroidal mode (corresponding to the white bars on Fig. A1a).

It was not be possible to observe toroidal dipolar response for the orthogonal polarization, i.e. when magnetic vector of the plane wave had been set parallel to the axes of the cylinders (Fig. A2). Although such polarization is shown to induce magnetization in a dielectric cylinder [53], the corresponding magnetic dipole moment is oriented parallel to the axis of the cylinder. In a cluster of 4 infinitely long cylinders such orientation cannot lead to the formation of a magnetic-field vortex characteristic to the toroidal mode (see insets to Fig. A2).

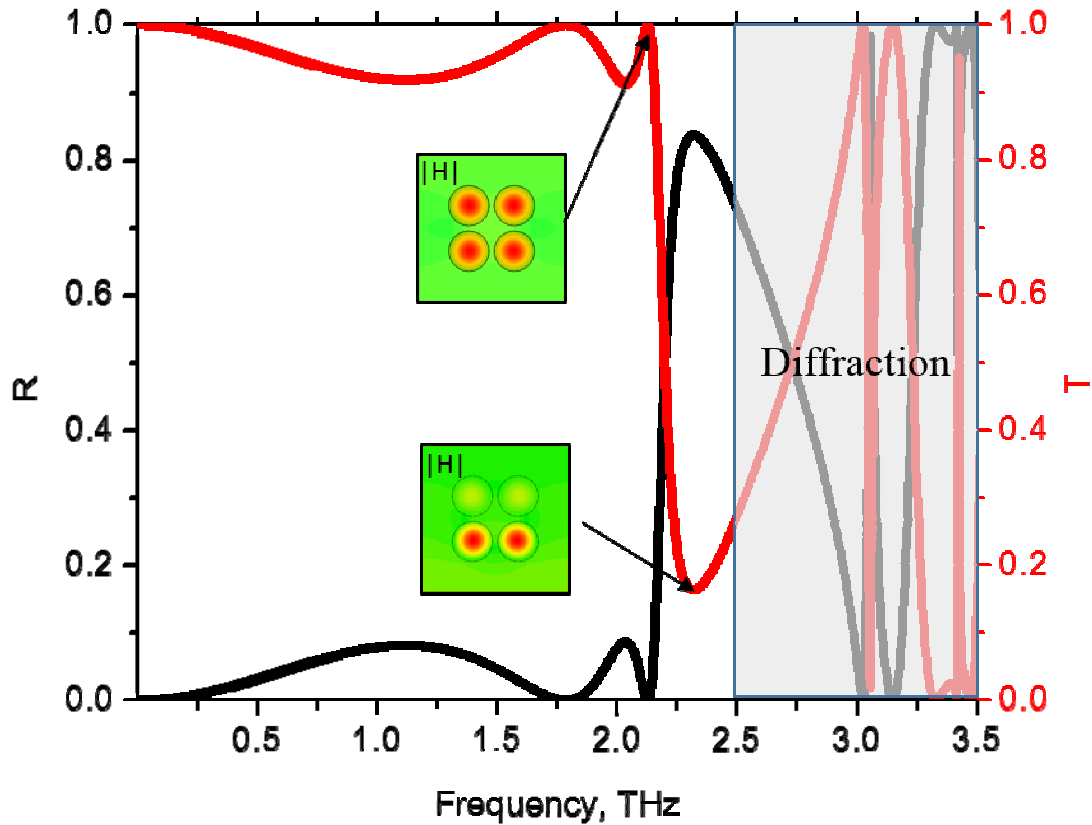

Figure A2. Transmission, T (red), and reflection, R (black), spectra of the metamaterial slab calculated with CST Microwave Studio for normally incident H-polarized plane waves (magnetic vector of the waves was set parallel to the axes of the cylinders). Insets show magnetic near-field distributions calculated at 2.13 and 2.32 THz.

We note in passing that the metamaterial response to normally incident plane waves also featured a very strong contribution from magnetic quadrupole $\mathbf{Q_m}$ near 1.95 THz (Fig. 3b). As in the case of toroidal dipolar excitation it emerges as the result of coupling of individual magnetic dipole modes induced in each of four cylinders (see Fig. A3). Contrary to the toroidal dipolar mode, the electric field of the magnetic quadrupole is completely expelled from the central region of the metamolecule. This presents an intriguing opportunity of using the metamaterial as a microscopic cloaking device, where the presence of an object placed in the center of the metamolecule may not be revealed.

(a)　　　　　　　　　　　　　　　　　(b)

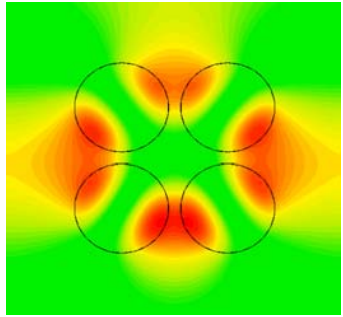
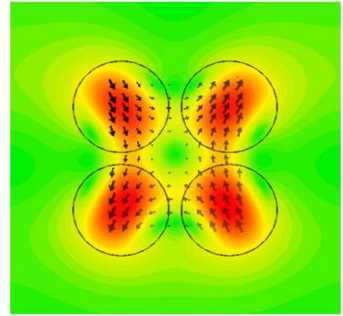

|$E_z$|　　　　　　　　　　　　　　　　　|$H$|

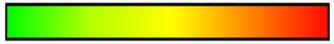
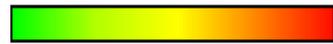

0　　　　　　　　　Max　　　　　　　　0　　　　　　　　　Max

Figure A3. (a) & (b) Calculated distributions of the absolute values of electric field ($z$-component) (panel b) and magnetic field (panel c) induced in the metamolecule at 1.95 THz. The slight asymmetry of the field distribution is attributed to the interference effects due finite size of the metamolecules in the direction of the wave propagation.